\documentclass[showpacs,twocolumn,prd,nofootinbib]{revtex4-1}
 \usepackage{color,graphicx,epsfig}
 \usepackage{amsmath,amssymb,dsfont,epsfig,graphicx,xcolor}
 \usepackage{ifpdf}
 \usepackage{amsmath}
 \usepackage{bm}
 \usepackage{color}
 \usepackage[english]{babel}
 \usepackage{graphicx}%
 \usepackage{amsfonts}%
 \usepackage{amssymb}
 \usepackage{braket}
 \usepackage{hyperref}
 \usepackage{epstopdf}
 \graphicspath{{Plots/}}

\definecolor{nicered}{rgb}{0.7,0.1,0.1}
\definecolor{nicegreen}{rgb}{0.1,0.5,0.1}
\definecolor{violet}{rgb}{0.7,0.3,0.3}
\hypersetup{colorlinks,citecolor= nicegreen,linkcolor= nicered}

\newcommand{\nc}{\newcommand}
\nc{\non}{\nonumber}
\nc{\hc}{\hbox {H.c.}}
\nc{\noi}{\noindent}
\nc{\barx}{\bar{x}}
\nc{\pbarn}{\;\hbox {pb}}
\nc{\fbarn}{\;\hbox {fb}}

\nc{\hsp}{\hspace{0.5cm}}
\nc{\lsp}{\hspace{1cm}}
\nc{\Lsp}{\hspace{2cm}}
\nc{\LLsp}{\lsp\lsp}
\nc{\lra}{\longrightarrow}
\nc{\p}{\prime}
\nc{\sgn}{\text{sgn}}
\nc{\ph}{\varphi}
\nc{\op}{{\cal O}}
\nc{\eq}{\text{Eq.~}}

\nc{\beq}{\begin{equation}}  \nc{\eeq}{\end{equation}}
\nc{\bea}{\begin{eqnarray}}  \nc{\eea}{\end{eqnarray}}
\nc{\baa}{\begin{array}}     \nc{\eaa}{\end{array}}
\nc{\bit}{\begin{itemize}}   \nc{\eit}{\end{itemize}}
\nc{\ben}{\begin{enumerate}} \nc{\een}{\end{enumerate}}
\nc{\bce}{\begin{center}}    \nc{\ece}{\end{center}}
\nc{\bpm}{\begin{pmatrix}}   \nc{\epm}{\end{pmatrix}}
\nc{\bvt}{\begin{verbatim}}  \nc{\evt}{\end{verbatim}}

\begin{document}

\def\LjubljanaFMF{Faculty of Mathematics and Physics, University of Ljubljana,
 Jadranska 19, 1000 Ljubljana, Slovenia }
\def\LjubljanaIJS{Jo\v zef Stefan Institute, Jamova 39, 1000 Ljubljana, Slovenia}

\title{Uncovering latent jet substructure}
 
\author{Barry~M.~Dillon}
\email[Electronic address:]{barry.dillon@ijs.si} 
\affiliation{\LjubljanaIJS}

\author{Darius~A.~Faroughy}
\email[Electronic address:]{darius.faroughy@ijs.si} 
\affiliation{\LjubljanaIJS}

\author{Jernej~F.~Kamenik}
\email[Electronic address:]{jernej.kamenik@cern.ch} 
\affiliation{\LjubljanaIJS}
\affiliation{\LjubljanaFMF}

\begin{abstract}

We apply techniques from Bayesian generative statistical modeling to uncover hidden features in jet substructure observables that discriminate between different a priori unknown underlying short distance physical processes in multi-jet events. In particular, we use a mixed membership model known as {\it Latent Dirichlet Allocation} to build a data-driven {\it unsupervised} top-quark tagger and $t\bar t$ event classifier. We compare our proposal to existing traditional and machine learning approaches to {top} jet tagging. 
Finally, employing a toy vector-scalar boson model as a benchmark, we demonstrate the potential for discovering New Physics signatures in multi-jet events in a model independent and unsupervised way. 

\end{abstract}

\maketitle

\section{Introduction}\label{intro}
The use of jet substructure techniques in studying large area jets has played an important role in  identifying hadronic decays of Higgs and electroweak gauge bosons in runs 1 and 2 of the LHC~\cite{Butterworth:2008iy,Butterworth:2002tt,Butterworth:2007ke,Cui:2010km}.
These techniques have also been used efficiently to tag jets arising from top quarks~\cite{Skiba:2007fw,Holdom:2007nw,Gerbush:2007fe,Kaplan:2008ie,Almeida:2008tp,Almeida:2008yp,Almeida:2010pa,Backovic:2012jk,Plehn:2009rk,Plehn:2010st,Soper:2012pb}.
In the last few years, machine learning (ML) tools have extended the application of jet substructure in tagging jets at the LHC~\cite{Larkoski:2017jix,Cogan:2014oua,Almeida:2015jua,deOliveira:2015xxd,Baldi:2016fql,Barnard:2016qma,Kasieczka:2017nvn,Butter:2017cot,Komiske:2016rsd,Louppe:2017ipp,Pearkes:2017hku,Datta:2017rhs,Datta:2017lxt,Fraser:2018ieu,Andreassen:2018apy,Macaluso:2018tck,Datta:2019ndh} through the use of Neural Networks (NNs) to process and `learn' from vast amounts of training data. Since these approaches rely on theoretical predictions for pure signal and background training data sets (typically through Monte Carlo (MC) generators), they  (a) are exposed to MC mismodeling of realistic events as reconstructed from real data and detectors; (b) require exact model knowledge of both expected signal and backgrounds. This limits their use in searches for a priori unknown new phenomena in LHC jet events.

There have been recent advances in unsupervised or semi-supervised ML techniques, based on NNs designed to be able to separate signal and background events in mixed samples, and could therefore be run directly on experimental data without the need for pure MC training samples, see e.g. refs.~\cite{Dery:2017fap,Metodiev:2017vrx,Komiske:2018oaa,Cohen:2017exh,Collins:2018epr,Collins:2019jip} and~\cite{Aguilar-Saavedra:2017rzt, Hajer:2018kqm, Heimel:2018mkt, Farina:2018fyg, Cerri:2018anq}.  They rely on categorizing and comparing datasets with different expected signal and background admixtures or identifying anomalous events inside large datasets. While these approaches ameliorate the model dependence of fully supervised ML, they are still potentially susceptible to correlated systematics (i.e. detector) effects and/or subject to large look-elsewhere effects. In addition, they generally work best when applied on very large datasets. Consequently their performance may suffer when looking for effects in tails of distributions.

In this Letter, we outline a new technique to classify jets and events {\it in situ} within a single mixed event sample, using tools developed in a branch of ML called generative statistical modeling~\cite{Deerwester90indexingby,Hofmann99probabilisticlatent}. Developed primarily to identify emergent themes in collections of documents, these models infer the hidden (or latent) structure of a document corpus using posterior Bayesian inference based on word and theme co-occurence~\cite{Nigam99textclassification}. Translated into the language jet physics, one assumes that observable jet substructure histograms ({\it words}) in events ({\it documents}) are generated by drawing from latent distributions ({\it themes}) of varying proportions. This allows to construct so called statistical mixed membership models of jet substructure.\footnote{Similar techniques have been used recently in a semi-supervised way to reconstruct `pure' quark and gluon jet observable distributions from mixed event samples~\cite{Metodiev:2018ftz, Komiske:2018vkc}.} Furthermore assuming that each event is a mixture of only few latent distributions and that within each of these only few histogram bins have high co-occurrence, such models can be solved using techniques of Latent Dirichlet Allocation (LDA)~\cite{Blei03latentdirichlet}. Finally, with a trained model at hand, one can define robust parametric jet and event classifiers by inferring on the latent distribution proportions in tested events.

In the following we first present the main ingredients of our proposal in more detail. Then we discuss two proof of principle implementations based on benchmark examples: an unsupervised top quark jet tagger and $t\bar t$ event classifier, as well as an unsupervised new physics (NP) search strategy able to identify boosted neutral scalar bosons decaying to pairs of $W$'s (previously studied in Refs.~\cite{Agashe:2018leo,Collins:2018epr,Collins:2019jip}). We compare them to existing conventional and ML approaches and also outline possible further improvements and future directions.\\

\section{Generative Bayesian Models of Jet Substructure}\label{TopicModels}
We start by considering the formation of a jet stemming from an initial hard seed, as a sequential combination of QCD showering (followed by fragmentation and hadronization) and possibly massive particle decays. Next we assume that some relevant information on this intertwined sequence of processes can be recovered by looking at the clustering history of a jet-clustering algorithm. This is in fact the basis for many conventional taggers of massive jets~\cite{Butterworth:2008iy, Kaplan:2008ie,Plehn:2009rk}. 

Within this very simplified picture of jet formation and observation we can draw interesting parallels to so called mixed membership models describing generation of documents in the context of text analysis~\cite{Blei03latentdirichlet}, or genotypes in population studies~\cite{Pritchard945}. In particular,  we assume that the observable distribution bins in a clustering profile are populated by drawing from a few {latent} distributions - {\it themes} - corresponding to different contributing physical processes. The likelihood of populating a certain distribution bin $o$, given a theme $t$ can then be described by a multinomial distribution $p(o|t,\beta)$ (a multi-category generalization of the binomial distribution, where the number of categories is given by the number of bins in the distribution and is parametrized by a set of parameters $\beta$).  In addition, we assume that the likelihood of a given theme contributing to any given event (and thus jet) $p(t|\omega)$ is also described by some multinomial distribution (parametrized by variables $\omega$), where the number of categories now corresponds to the number of themes. The $\omega$'s themselves are drawn from a probability distribution $p(\omega|\alpha)$, reflecting the theme proportions in the dataset and parametrized by the {\it hyperparameter} $\alpha$.  In this picture the themes ($\beta$) as well as theme proportions ($\omega$) are hidden variables reflecting the thematic structure of the studied event sample. 
With a given model, the probability that a certain event or jet distribution bin is populated can be written as a compact expression in terms of the latent variables. For example, the likelihood of generating a jet represented in terms of observables $j = (o_1,o_2,\ldots,o_n)$  is just
\beq \label{LDAref}
p(j|\alpha,\beta)=\int_\omega p(\omega|\alpha)\prod_{o\in j}\left(\sum_{t} p(t|\omega)p(o | t,\beta)\right)d\omega\,.
\eeq
Statistical models defined in this way are generative in that given the latent variables (themes and theme proportions) the best model will be the one that best reproduces a set of jets or events, i.e. has the best generative power.
Therefore, the task of finding the latent variables from a set of training events is specifically to invert the above expression and use the set of events to find the best fit for the latent variables.
This can in fact be done using posterior Bayesian inference, i.e.
\beq
p(a | x) \propto p(x | a) * p(a)\,,
\eeq
where $p(x | a)$ is the likelihood of observing $x$ given a latent variable $a$, while $p(a )$ and $p(a | x)$ are the prior and posterior distributions of the latent variable itself. The main insight here is that $p(\omega|\alpha)$ in Eq.~\eqref{LDAref} is a conjugate prior to the multinomial likelihood $p(t|\omega)$ and thus forms the multi category generalization of the beta distribution - the Dirichlet distribution. The model is thus called LDA and can be solved approximately ({\it trained}) in an iterative manner using variational inference ~\cite{Blei03latentdirichlet,Hofmann99probabilisticlatent} or Gibbs sampling~\cite{Griffiths5228}. 

{The generative model defined by Eq.~\eqref{LDAref} does not include the conditional probabilities $p(o_i|o_{i-1})$ describing the ordering present in the (binary) clustering tree of the jet (or correspondingly in a Markov chain Monte Carlo jet generator). Therefore, the jet observables at each clustering step are assumed to be ``conditionally independent"~\cite{Blei03latentdirichlet}, i.e. they only depend conditionally on the same latent distributions ($\beta$, $\omega$) of the model. This is reminiscent with the {\it bag-of-words} assumption widely used in probabilistic text modelling where the semantic structure relating different words in the vocabulary is completely neglected in the generative process for documents. While this simplifying assumption, of neglecting the clustering order information in jets, forbids us to use the probabilistic model \eqref{LDAref} as a reliable jet or event generator\footnote{In the same way most generative text models can not be used as reliable document generators.}, it still comes in useful for jet or event classification tasks. As we show below, the LDA generative model is flexible enough to capture hidden features in the jet clustering history, in particular, features produced by the decay chains of massive resonances.

Formally, a trained LDA model consists of the latent variables inferred from the training data and the probabilistic generative model used in constructing Eq.~\eqref{LDAref}. In order to classify jets or events, we can perform statistical inference on the test sample. Once the LDA model is trained, the theme proportions ($\omega_t(j)$) present in each new jet $j$ (or event) can be estimated by maximizing the likelihood function for $j$ while keeping the theme distributions ($\beta$) fixed. As a result, each jet is described by a mixture of themes with corresponding weights $\omega_t (j)$ that can be directly used for classification. Since the extracted mixtures satisfy $\sum_t \omega_t (j) = 1$ and here we are focusing on only two themes (i.e. $t=0,1$) it suffices to choose just one of the weights to describe the jet. In this case, we define a simple classifier $h(j) = \omega_1(j)$ based on the proportion of one of the themes in the jet (or event). 

Alternatively, one can directly use the latent themes $p(o|t_{1,2})$ discovered by the LDA algorithm and compute the likelihood ratio 
$\mathcal{L}(j)={p(j|t_1)}/{p(j|t_2)}\,$
for every new jet $j$ (or event) in a test sample, and use it as the classifier. 
While the likelihood ratio is known to be the optimal classifier given exact knowledge of pure distributions~\cite{NeymanPearson}, it has been shown recently, that it remains optimal even for mixed distributions of a-priori unknown but different mixture proportions~\cite{Metodiev:2017vrx}. Thus, $\mathcal{L}(j)$ is an optimal LDA classifier in the limit that the extracted themes correspond to pure distributions and the LDA model reduces to a simple mixture model. In general however, this will not be the case and we have checked explicitly that the inference and $\mathcal{L}(j)$ based classifiers based on LDA perform comparably. In the remainder of the paper we only present results based on the inference classifier $h(j)$.}\\

\section{Unsupervised top tagger}\label{tttagger}
Our first proof of principle example is a tagger discriminating between boosted hadronically decaying top quarks and QCD jets. Working with a single mixed ($t\bar t$ and QCD) multi-jet event sample we first need to construct the relevant jet substructure observable histograms ($o$). We do this by clustering the jets in an event using the Cambridge-Aachen (CA)~\cite{Dokshitzer:1997in, Wobisch:1998wt} algorithm with a large radius $R$. We then proceed to uncluster the jets by reversing each step in the clustering, iteratively separating each (sub)jet into two objects $j_0 \to j_1 j_2$. Ordering the subjets by their invariant mass $m_{j_1}>m_{j_2}$ (and following the standard approach of refs.~\cite{Butterworth:2007ke,Butterworth:2008iy}), we define the relevant clustering observables at each clustering step as
\begin{align}
o_{j_0} = \Big\{ m_{j_{0}}\,, && \frac{m_{j_1}}{m_{j_0}}\,, && \frac{m_{j_2}}{m_{j_1}}\,, && \frac{{\rm min} (p^2_{T,1} , p^2_{T,2})}{m^2_{j_0}} \Delta R^2_{1,2} \Big\} \,,
\end{align}
where $p_{T,i}$ is the transverse momentum of a given object $j_i$ and $\Delta R_{1,2}^2 = (\phi_{1}-\phi_{2})^2 + (\eta_{1} -\eta_{2})^2$ is the so called planar distance between $j_1$ and $j_2$ ($\phi_i$ and $\eta_i$ being the azimuthal angle and pseudorapidity of $j_i$, respectively).
The declustering step is then iteratively repeated on both $j_{1,2}$. The procedure is terminated once $m_{j_0}<m_{\rm min}$, where $m_{\rm min}$ is an algorithm parameter, which we choose to lie below the lowest massive resonance state of interest. In the case of the top tagger, we fix $m_{\rm min} =30\ {\rm GeV}  \ll m_W$, but have checked that lowering this threshold by a factor of a few does not significantly affect the results. 
The output of such a procedure is a (typically a rather sparse) four-dimensional histogram of $o_{j}$ which can be defined either {\it per jet} or even {\it per event}. After mapping individual histogram bins into {\it words}, we feed individual jets or events as {\it documents} into an LDA implementation using the software package \texttt{Gensim}~\cite{rehurek_lrec,Hoffman:2010:OLL:2997189.2997285}, fixing the number of themes to two ($\omega_{0,1}$). Further technical details of the required binning and mapping of data onto (one-dimensional) text vocabularies compatible with \texttt{Gensim}, as well as a detailed analysis of the convergence of the algorithm when applied on sparse jet substructure data will be presented elsewhere~\cite{longPaper}. 
Here we only focus on the consistency and stability of the resulting trained models. For this purpose we use the $k$-folding method with $k=10$.
This involves splitting the training data into $k$ different mutually exclusive blocks and then running the training $k$ times on event samples built from $k\!-\!1$ blocks, with the combination changing on each training run. The performance of the tagger is tested on events or jets from the remaining block. 

In order to evaluate the performance of the tagger and compare it to existing methods, we construct a receiver operating characteristic (ROC) curve for our tagger. This is the only step where one needs to rely on access to pure samples (either MC generated or pre-tagged in some other way using observables orthogonal to $o_j$). In particular, we construct the ROC curve by performing the classification on such pure samples while continuously varying the threshold of the theme proportion defining the classifier $h(j)$. 
This is done for all $k$ sets of results and we calculate the median mis-tag rate ($\varepsilon_b$) for each signal efficiency ($\varepsilon_s$), as well as the mean absolute deviation of the mis-tag rate to evaluate the stability and consistency of the tagger.

Our training samples for the QCD di-jet background and the (hadronic) $t\bar{t}$ signal both consist of $\sim\!84,000$ $13$~TeV $pp$ collision events, where the final state particles are clustered into $R\!=\!1.5$ CA jets with $p_T$ in the range $[350,450]$~GeV. The samples are generated using {\tt aMC@NLO~2.6.1}~\cite{Alwall:2014hca} interfaced with {\tt Pythia~8.2}~\cite{Sjostrand:2007gs} for showering and hadronization, while jet clustering is performed using {\tt FastJet~3.2.0}~\cite{Cacciari:2011ma}. Note that no grooming is performed on the jets. We have also checked explicitly that applying jet (sub)cluster energy smearing consistent with the parametric fast detector simulation of ATLAS implemented in {\tt Delphes 3.4.1}~\cite{deFavereau:2013fsa} has no significant effect on our results.

We train the top tagger on four test cases: supervised, and unsupervised mixed samples with $S/B=1,~1/9,~1/99$.
In the supervised case we collapse the pure samples into single documents such that they are processed by the algorithm in a single block, essentially providing the labelling of the data required in supervised algorithms.
For the different $S/B$ ratios each jet or event is represented by a single document. However, we inform the tagger to search for certain $S/B$ ratios by setting the hyperparameters of the Dirichlet distribution accordingly, i.e. $\alpha=[0.5,0.5]$, $[0.9,0.1]$, and $[0.99,0.01]$.  Note that these may not be the optimal choices, but they are based on the intuition from the values of $S/B$ and give a useful parameterization to demonstrate the performance of the algorithm. 
{ We also stress that $\mathcal{O}(1)$ variations in $\alpha$ have only a small effect on the performance of the algorithm provided that the hierarchy in the elements of $\alpha$ approximately reflect the $S/B$ ratio, and that the elements are smaller than one.}
More details on the dependence of the algorithm on these hyperparameters, and how to determine their optimal values without prior knowledge of the $S/B$ ratios, will be presented elsewhere~\cite{longPaper}.

\begin{figure}[t!]
\centering
\includegraphics[width=8.5cm]{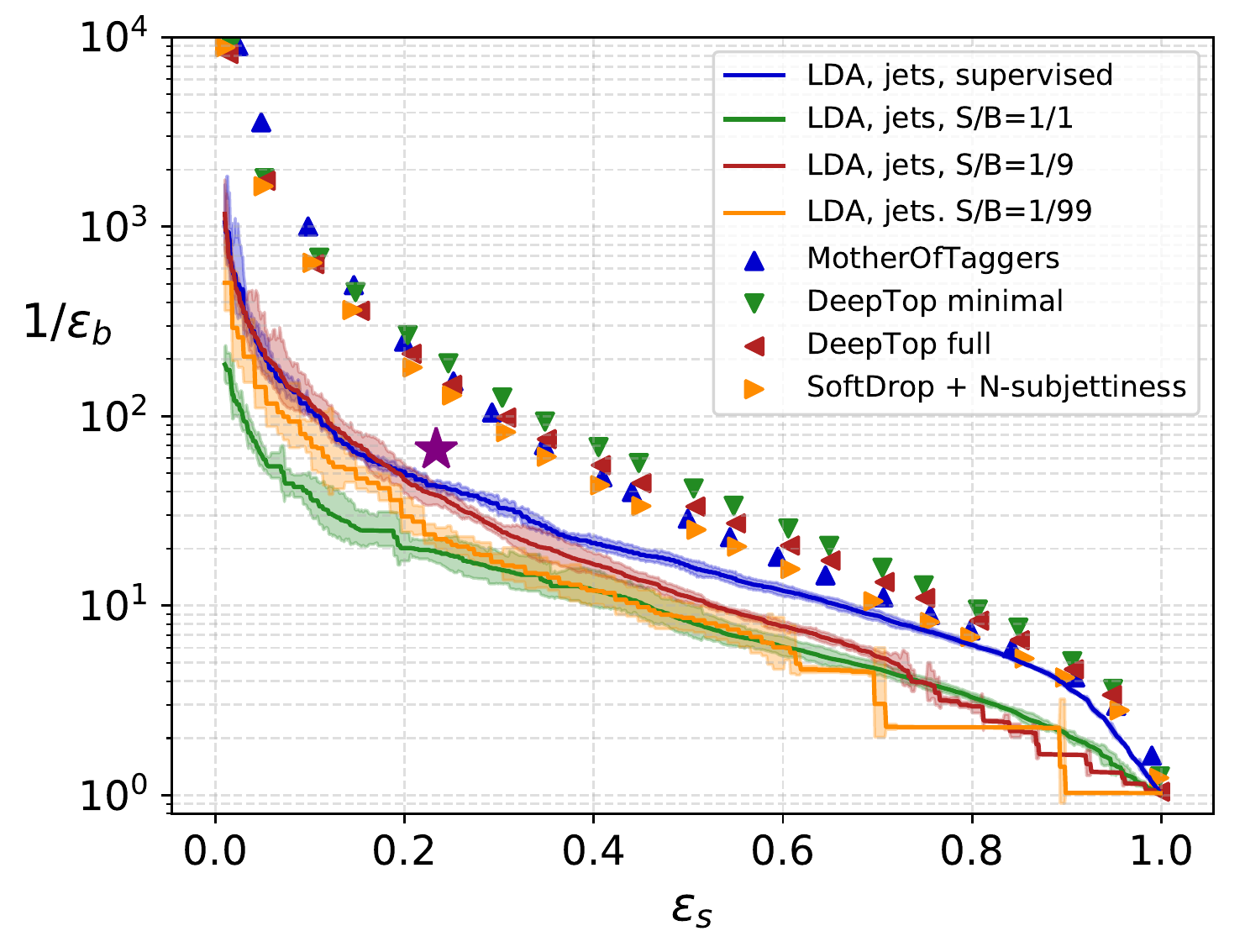}\\
\includegraphics[width=8.5cm]{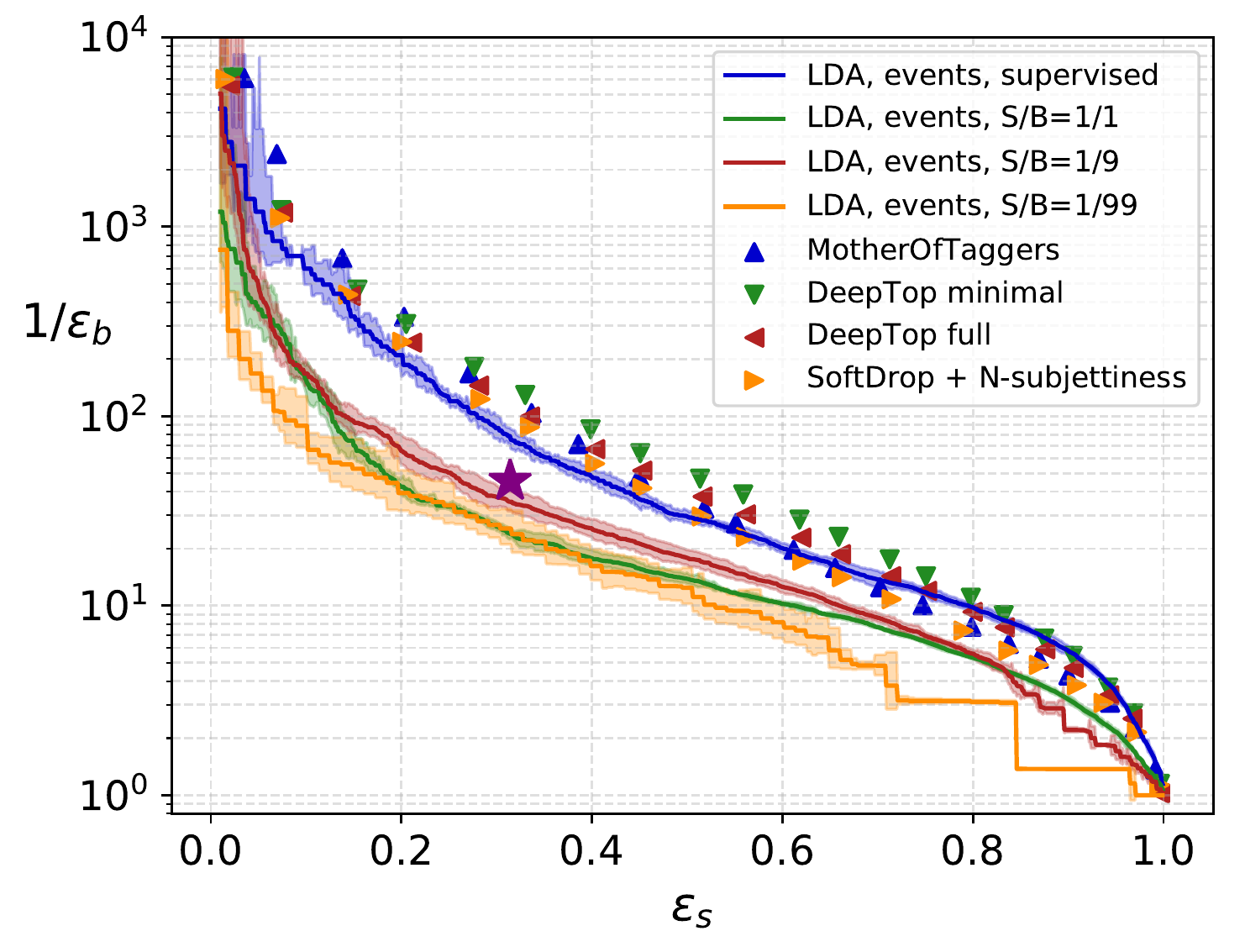}
\caption{(Upper plot) ROC curves for the LDA top jet taggers compared to the DeepTop tagger~\cite{Kasieczka:2017nvn,Butter:2017cot} (colored triangles) for events with fat-jets satisfying $p_T\in[350,450]$~GeV. The purple star represents the default JH top tagger~\cite{Kaplan:2008ie} reference point. (Lower plot) ROC curves for the $t\bar t$ LDA event classifiers compared to the classifiers from the DeepTop (colored triangles) and the JH top tagger (purple star). In both plots the shaded bands represent the mean-average-deviation extracted from the $k$-folding procedure. See text for details.}
\label{ttTaggerROCjets}
\end{figure}

\begin{figure}[t!]
\centering
\includegraphics[width=8.5cm]{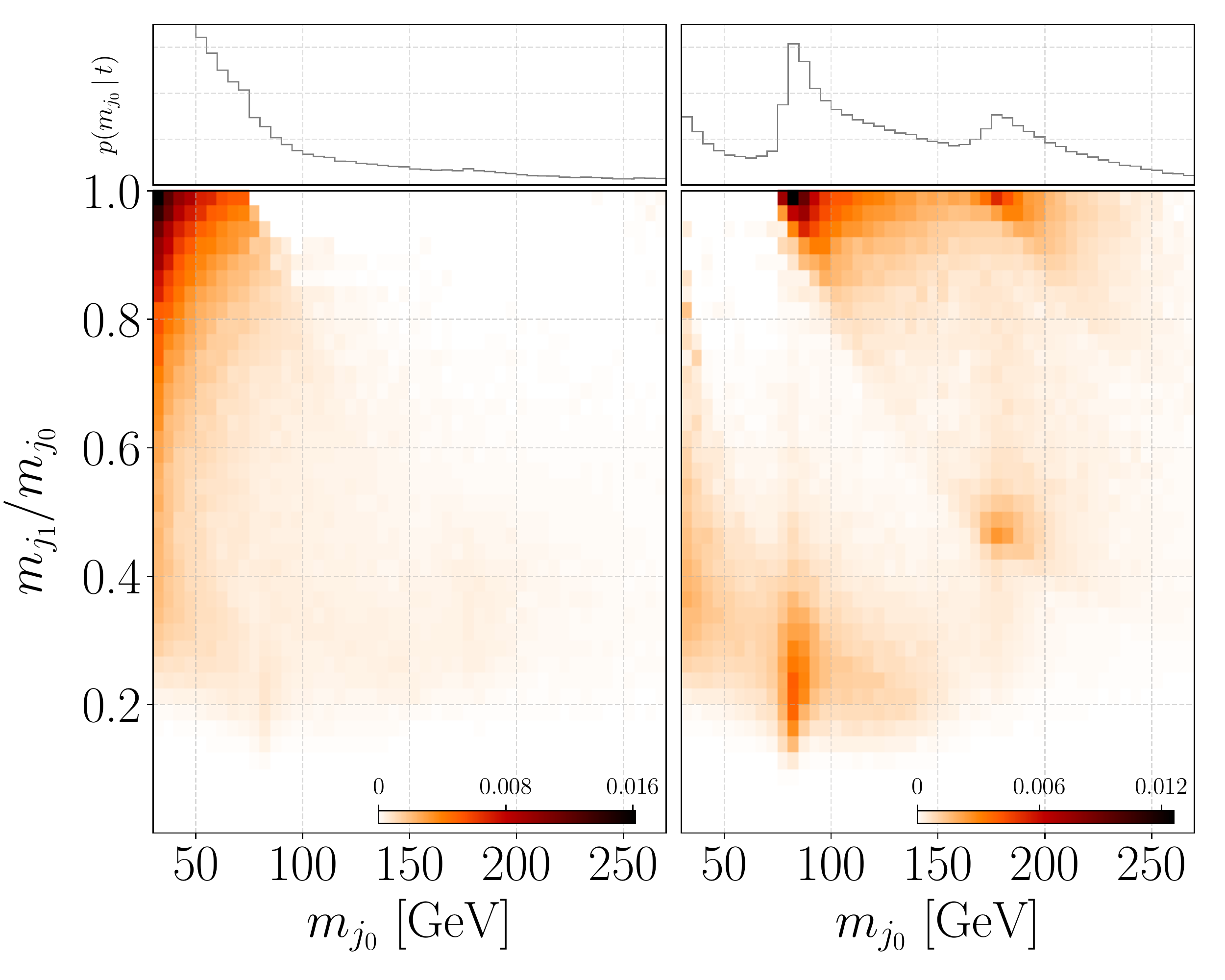}
\caption{2D projected probability distributions (in the plane of $m_{j_0}$ and $m_{j_1}/m_{j_0}$ ) of the two latent themes discovered in mixed  ($S/B=1$) QCD and $t\bar t$ event samples with fat-jets satisfying $p_T\in[350,450]$~GeV.}
\label{ttTaggerHMjets}
\end{figure}

In Fig. \ref{ttTaggerROCjets} (upper panel) we plot the ROC curves for our top jet taggers, where separate documents are represented by individual jets, and compare these to various supervised taggers in the literature~\cite{Kaplan:2008ie,Kasieczka:2017nvn,Butter:2017cot}.
We see that the taggers perform well and with relatively small variance, with the supervised tagger performing the best.
An interesting observation is that at high background rejection rates ($1/\epsilon_b \gg \mathcal O({\rm few})$) the taggers trained on smaller $S/B$ perform slightly better than the tagger trained on the $S/B=1$ sample, although the differences are comparable to the estimated uncertainties.
This is essentially because the algorithm is designed to discern features in the jet substructure, which are subsequnetly used to tag jets and events.
In the supervised and $S/B=1$ case the algorithm discovers features in top jets both near $m_{j_0}\sim m_t$ and $m_{j_0}\sim m_W$ (see the right plot in Fig.~\ref{ttTaggerHMjets}), while in the lower $S/B$ cases the algorithm is only able to identify $m_{j_0}\sim m_t$ as relevant.
On the other hand, lower $m_{j_0}$ regions generically feature more prominently in QCD jets (see left plot in Fig.~\ref{ttTaggerHMjets}). 
Thus, while a very accurate determination of the features near $m_{j_0}\sim m_W$ in the supervised case helps the performance of the tagging algorithm, the worse resolution in the unsupervised $S/B=1$ case leads to worse tagging performance compared to lower $S/B$ examples.
We see that the performance of the unsupervised taggers is comparable to the original JH top tagger~\cite{Kaplan:2008ie}, although it falls short in comparison to the others. We note that the observables we use mostly match those used in the JH top tagger, hence the similar performance is indeed encouraging.

In Fig. \ref{ttTaggerROCjets} (lower panel) we plot the ROC curves for our $t\bar{t}$ event classifiers, where a single document now contains all jets within the selected $p_T$ region in an event, and again compare these to the top jet taggers in the literature.
To make the comparison with other taggers fair, we re-scale those results by defining an event tagging efficiency $(\epsilon_e)$ in terms of the jet tagging efficiency $(\epsilon_j)$ and the fraction of events in our pure samples with one $(f_1)$ and two $(f_2)$ jets passing the selection cuts\footnote{We have checked that the fractions of events with zero or more than two jets passing the selection cuts are negligible.},
$\epsilon_e=(2\epsilon_j-\epsilon_j^2)f_2+\epsilon_j f_1$.
This means in practice that tagging an event as $t\bar{t}$ requires at least one jet in the event to be tagged as a top jet.
The ROC curves do not change significantly under this re-scaling, instead the points move along a trajectory towards higher efficiencies approximately equal to that of the ROC curve for jet tagging.
We see again that the classifier performs very well in all cases, performing as well as the JH top tagger even for low $S/B$.

We observe that the LDA algorithm performs relatively better when characterizing and tagging events than jets, mainly due to the larger amount of substructure ({\it words}) in each document. With more data per document it is easier for the algorithm to identify co-occurrences between the different features shared by jets in the same  event. For this reason it is also easier for the trained model to infer the correct thematic structure from events, than from jets.

The themes discovered by the unsupervised training algorithm contain valuable information about the substructure of the events or jets.
In Fig. \ref{ttTaggerHMjets} we plot the substructure probability distributions of the two themes discovered by the top jet tagger (with $S/B=1$) projected onto the plane of $m_{j_0}$ and $m_{j_1}/m_{j_0}$.
We observe that while the distribution on the left-hand side plot (the ``QCD" theme) is fairly unremarkable (mostly monotonic and smooth) and peaks towards ($m_j \to 0$, $m_{j_1}/m_{j_0} \to 1$), the theme on the right-hand side plot (the ``$t\bar{t}$" theme) clearly exhibits a heavily weighted feature at both $m_{j_0}\sim m_t$ and $m_{j_0}\sim m_W$, even identifying the $W$ subjet arising from the decay of the top quark within the jet resulting in a mass drop of $m_{j_1}/m_{j_0} \sim m_W / m_t \simeq 0.45$.
On the other hand, the broad $m_{j_1}/m_{j_0} \sim 0.2 \gtrsim 0$ feature at $m_{j_0}\sim m_W$ 
is expected due to the fact that the mass drop is defined with the heaviest daughter subjet in the numerator thus skewing the $m_{j_1}/m_{j_0}$ distribution away from zero.\\

\section{Unsupervised NP search}\label{nptagger}
As a second example, we consider a NP model~\cite{Agashe:2016rle,Agashe:2017wss} containing a heavy $W'$ boson plus a heavy scalar $\phi$. Signal events thus consist of resonant $W'$ production (at $m_{W^\prime}=3$ TeV), followed by $W' \to W \phi$ decays  (where we choose $m_\phi=400$ GeV$ \ll m_{W'}$  such that both the $W$ and the $\phi$ coming from $W'$ decays are boosted). Finally, the scalar further decays as $\phi \to W^+W^-$.
Using the same event generation, jet clustering/de-clustering procedure, observable basis $o_j$, and the same LDA tagging algorithm as before, we apply our procedure to the all-hadronic final state of this NP process in a region dominated by QCD background. The same model has been previously studied using the unsupervised ML approach called {\it classification without labels} (CWoLa)~\cite{Collins:2018epr,Collins:2019jip}. It is based on mixed sample classification using phase space regions with vastly different S/B ratios processed by deep NNs. In order to quantitatively compare our results to CWoLa, our signal and background event samples mirror directly those in Ref.~\cite{Collins:2019jip}.
In particular, we consider just the signal region, $2730\leq m_{jj}\leq3189$~GeV, and cut jets with $p_T$ below $400$ GeV.
The $30$~GeV cut on the subjet invariant mass is also applied, just as in the top tagger case.
After the selection cuts we work with $\sim\!60,000$ events in both the signal and background samples.
We train three different taggers;  a supervised tagger, and two taggers with $S/B=1.1\times 10^{-2}$ and $5.8\times 10^{-3}$.
The Dirichlet hyperparameters $\alpha$ are chosen in the same way as in the previous section, i.e. $\alpha=[0.5,0.5]$, $[0.989,0.011]$, and $[0.942,0.058]$. To evaluate the robustness of the taggers we again employ the $k$-folding procedure with $k=10$.

\begin{figure}[t!]
\centering
\includegraphics[width=8.5cm]{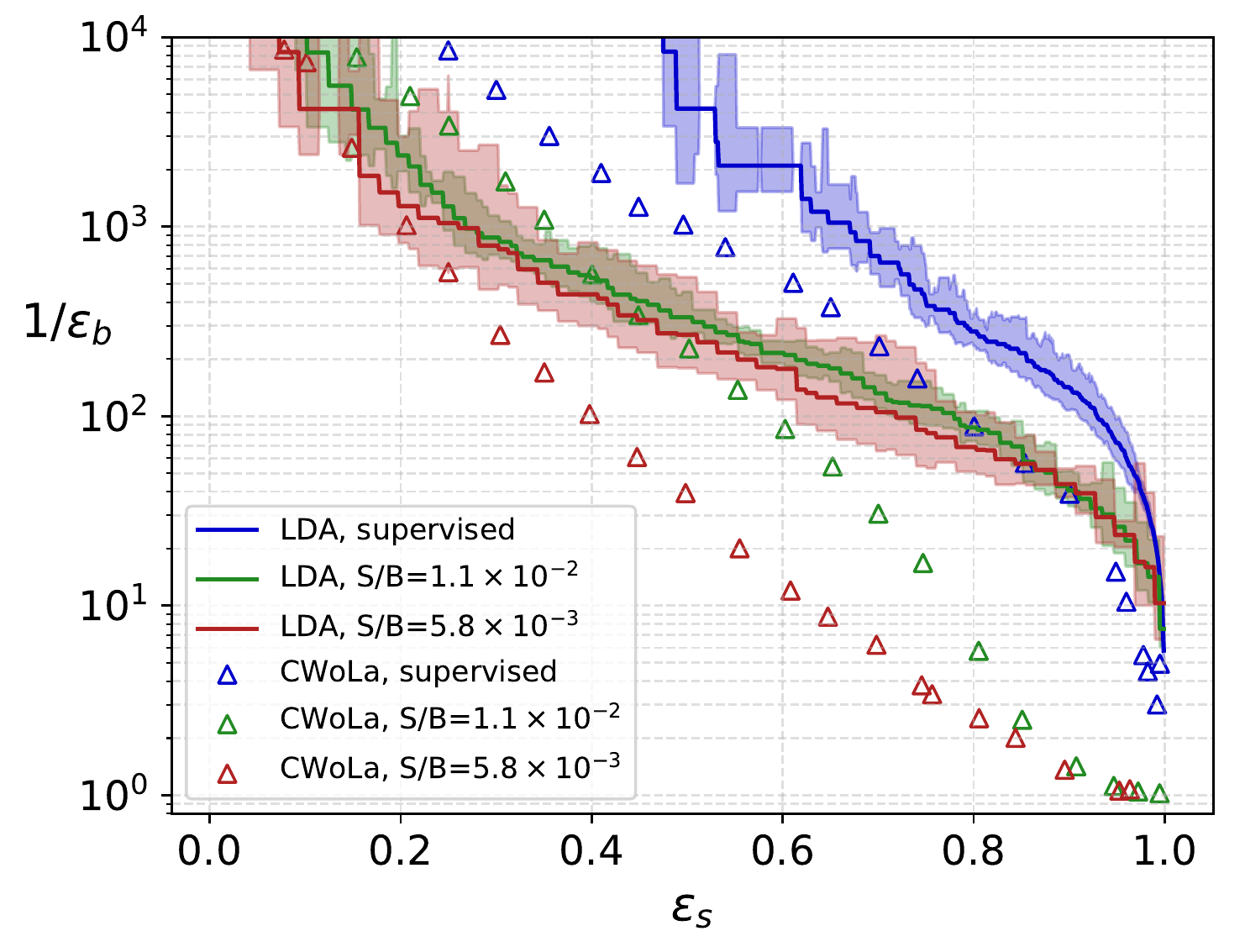}
\includegraphics[width=8.5cm]{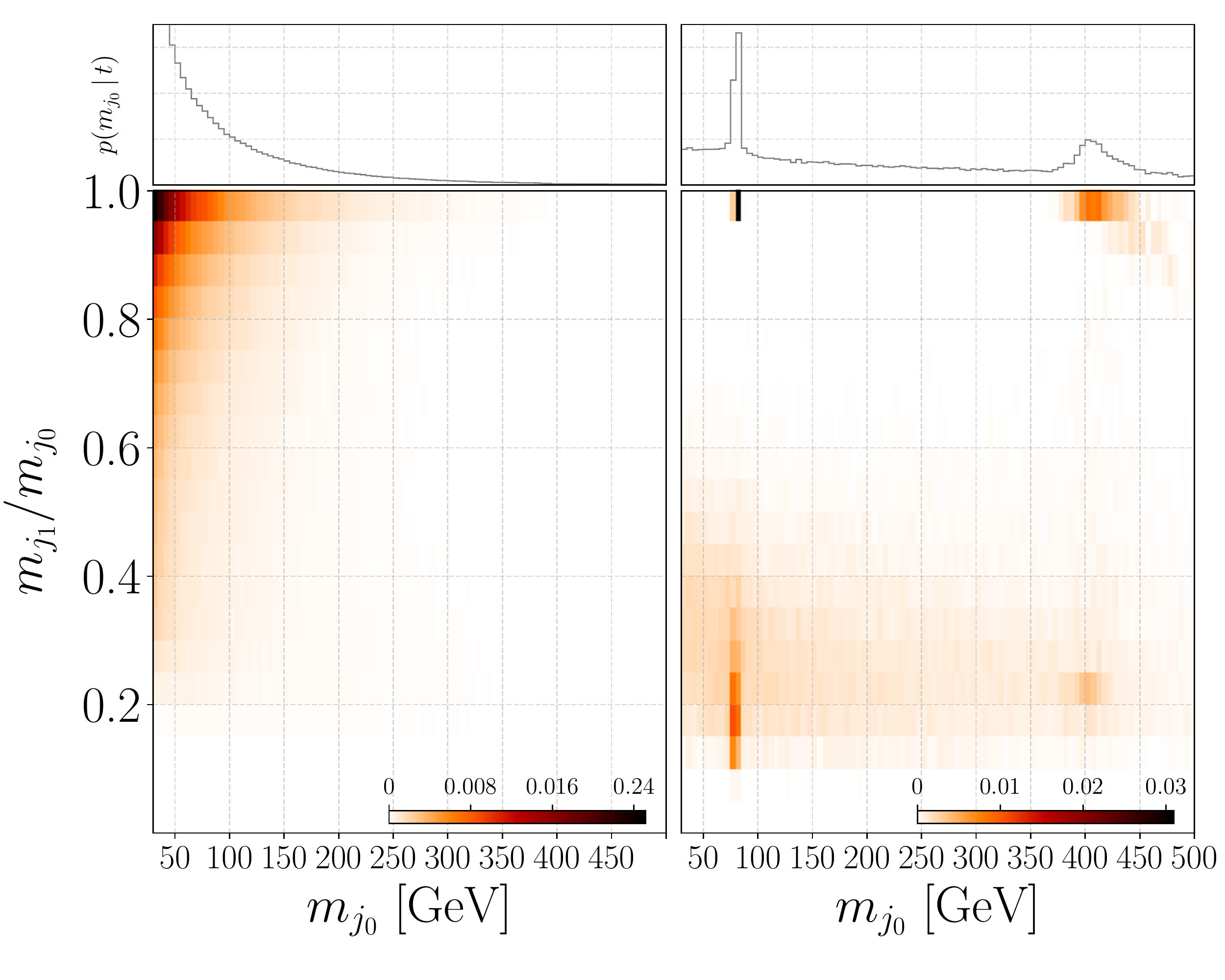}
\caption{(Upper plot) ROC curves comparing the performance of the LDA event classifier to CWoLa~\cite{Collins:2019jip}.
(Lower plot) 2D projected probability distributions (in the plane of $m_{j_0}$ and $m_{j_1} /m_{j_0}$ ) of the two latent themes discovered in mixed ($S/B = 1.1 \times 10^{-2}$) QCD and $W'$ event samples with invariant mass $2730\leq m_{jj}\leq3189$~GeV with fat-jets satisfying $p_T > 400$~GeV.
\label{WpTagger}}
\end{figure}

In the upper plot of Fig.~\ref{WpTagger} we show the ROC curves for our taggers and compare the results to those from CWoLa~\cite{Collins:2019jip}.
We see that in most of the parameter space the LDA-based tagger outperforms the CWoLa tagger, most notably at high signal efficiencies.

In the lower plot of Fig. \ref{WpTagger} we also show the probability distributions of the discovered themes in the plane of $m_{j_0}$ and $m_{j_1} /m_{j_0}$ for the LDA model  trained on event samples with $S/B=1.1\times 10^{-2}$.
Features in the subjet mass at $m_{j_0}\sim m_W$ and at $m_{j_0}\sim m_\phi$ are clearly discernible in one of the themes (the ``$ \phi W$" theme), as well as mass drops related to the decays of the heavy scalar and the $W$ bosons.\\

\section{Conclusions}\label{conclusions}
We have demonstrated a new unsupervised ML technique for disentangling signal and background events in mixed samples by identifying features in jet substructure observables that differentiate between the two.  
To do so we have mapped jet substructure distributions onto a LDA model, a generative probabilistic model ({\em mixed membership model}) widely used in Bayesian statistics approaches to unsupervised ML. 
Assuming that the kinematic observable distributions within jets or events are sampled from a fixed set of (latent) themes, LDA can learn the thematic structure that most likely generated the observed data (the later being either in the form of reconstructed real LHC events or un-labeled MC-generated samples). Furthermore, we have shown that the learned structure from a two-theme LDA model can be used to build unsupervised jet taggers or event classifiers that efficiently discriminate between signal and background in previously unseen data.

As a first example we have trained a two-theme LDA model on MC-generated event samples consisting of different mixtures of $pp\to t\bar t$ and QCD di-jet events. Our results show that the top-jet taggers and $t\bar t$ event classifiers constructed from the discovered themes have a very good discrimination power when applied to previously unseen pure samples, even if trained on data with $S/B$ ratios as low as $1\%$. Our results are in some cases comparable even with fully supervised taggers in the literature. In addition we have explored the viability of LDA discovering NP phenomena in multi-jet events. Using a benchmark NP (vector $W'$ - scalar $\phi$) model we have studied $pp\rightarrow W'\rightarrow \phi W\rightarrow WWW$ with hadronically decaying $W$ bosons and a (boosted) new scalar $\phi$ with mass $m_\phi\ll m_{W'}$. The resulting LDA event classifiers from training samples with $S/B$ as low as a few per-mille, when applied to pure samples, produce excellent signal efficiencies and QCD rejection rates that can outperform other existing approaches. 

Besides being a fully unsupervised ML technique, one advantage of performing LDA on jet clustering history observables, is the possibility of interpreting the thematic structure discovered by the model from the data. In both examples presented here, the features in the probability distributions over the kinematical observables of the two uncovered themes match to a high degree the expected features of the underlying hard processes - hadronic decays of top-quarks (or $\phi\to W^+ W^-$) and the QCD background, respectively, allowing for an intuitive and physical understanding of the high tagging performance as demonstrated by the ROC curves. 

The analysis presented here is a first exploration of what can be achieved when applying probabilistic mixed membership models to high-energy collider data. For example, with the addition of more jet substructure observables the discriminating power of the LDA classifiers could be further optimized and increased. Furthermore, relaxing the fixed number of themes of the LDA model applied to mixed event samples could allow to classify multiple backgrounds together with the signal. In future work we will also detail how these techniques can be employed as part of a broad search strategy for new phenomena in multi-jet invariant mass spectra with the aim of performing unsupervised data-driven searches for NP at high $p_T$. 

\begin{acknowledgments}

We thank Jasna Urban\v ci\v c, Erik Novak and Klemen Kenda for initial involvement in the project as well as Jack Collins for generously providing the $W'-\phi$ NP model implementation for use in aMC@NLO. We also thank C\'esar A. Ojeda and Bryan Zaldivar for useful discussions. DAF is supported by the Young Researchers Programme of the Slovenian Research Agency under the grant No. 37468. JFK and BMD acknowledge the financial support from the Slovenian Research Agency (research core funding No. P1-0035 and J1-8137).

\end{acknowledgments}

\bibliographystyle{elsarticle-num}
\bibliography{current}

\end{document}